\documentclass[letter]{jpsj3} 
\title{First-Principles Momentum-Dependent Local Ansatz Wavefunction and
Momentum Distribution Function Bands of Iron}

\author{
Yoshiro \textsc{Kakehashi}\thanks{E-mail address:
yok@sci.u-ryukyu.ac.jp, be published in J. Phys. Soc. Jpn. {\bf 85} 
(2016).} and Sumal \textsc{Chandra}
}

\inst{
Department of Physics and Earth Sciences,
Faculty of Science, \\
University of the Ryukyus, \\
1 Senbaru, Nishihara, Okinawa 903-0213, Japan
}

\abst{
We have developed a first-principles local ansatz wavefunction
approach with momentum-dependent variational parameters on 
the basis of the tight-binding LDA+U Hamiltonian.  The theory goes
beyond the first-principles Gutzwiller approach and quantitatively 
describes correlated electron systems.  Using the theory, 
we find that the momentum distribution function (MDF) bands 
of paramagnetic bcc Fe along high-symmetry lines show a large 
deviation from the
Fermi-Dirac function for the $d$ electrons with $e_{g}$ symmetry and
yield the momentum-dependent mass enhancement factors.  The calculated
average mass enhancement $m^{\ast}/m = 1.65$ is
consistent with low-temperature specific heat data as well as 
recent angle-resolved photoemission spectroscopy (ARPES) data.
}

\kword{first-principles momentum-dependent local ansatz,
Gutzwiller wavefunction, density functional theory,
momentum distribution function, mass enhancement, electron 
correlations, ARPES
}

\begin{document}
\maketitle

%
%
%

The quantitative description of correlated electrons in solids has been one
of the central issues in condensed matter physics.
Band theory has been developed towards such a quantitative
description by taking into account electron-electron 
interactions.
Density functional theory (DFT) 
is the state-of-the-art theory along this line.  
The DFT is based on the Hohenberg-Kohn theorem~\cite{kohn64} 
and the Kohn-Sham method~\cite{kohn65}. 
The local density approximation (LDA) and
the generalized gradient approximation (GGA) to the exchange 
correlation potential led to the development of the DFT as a 
quantitative theory.

Although the DFT has been successful for the quantitative understanding of
metals and band insulators, it has failed to provide a quantitative 
description of more correlated electron systems such as 
$\epsilon$-Fe~\cite{pour14}, Fe pnictides~\cite{imada10}, 
cuprates~\cite{fulde12}, as well as heavy-fermion 
systems~\cite{fulde12}.
Moreover, the DFT does not describe the physical quantities such as 
charge and spin fluctuations, the momentum distribution function (MDF), 
as well as related excitations.  

An alternative approach to the quantitative description of
correlated electrons is to
construct a first-principles effective tight-binding Hamiltonian that is 
compatible with the standard many-body theories and combine the
Hamiltonian with them.  The dynamical mean field theory (DMFT) combined
with the LDA+U Hamiltonian~\cite{kotliar06,anisimov10}, 
which is equivalent to the first-principles
dynamical coherent potential approximation (DCPA) that we 
developed~\cite{kake11,kake13}, is such an approach using 
the Green function technique and an effective medium.
The theory allows us to study strongly correlated electrons within a
single-site approximation.

The third approach towards the quantitative description is to combine
the first-principles LDA+U Hamiltonian with the Gutzwiller variational
theory~\cite{bune00,bune12}.  
The Gutzwiller wavefunction describes the ground state of
correlated electrons by means of the projection operators which extract
the atomic states on each atom from the Hartree-Fock state.  
The first-principles Gutzwiller theory can resolve a small energy
difference between the states, which cannot be achieved by the 
LDA+DMFT~\cite{schick12}.  Thus, the theory 
has been applied to many systems and has clarified the physics of
electron correlations~\cite{bune12}.  
More recently,  generalized Gutzwiller
density functional theory (GDFT)~\cite{bune12,schick12,deng09}, 
in which the wavefunction is used as a reference system 
to the DFT energy, has even been developed for more
accurate self-consistent calculations. 

The Gutzwiller wavefunction, however, does not reduce to 
the second-order perturbation theory in the weakly correlated 
limit~\cite{kake08}.
Thus, it does not quantitatively describe the properties in the weakly
correlated regime.  

 In order to overcome the difficulty in the Gutzwiller-type  
variational theories, 
we have recently proposed a momentum-dependent local ansatz (MLA) 
wavefunction~\cite{kake08,pat11,pat13}.  
The MLA is a generalization of the local ansatz approach
(LA) by Fulde and Stollhoff~\cite{stoll77,stoll81}, and 
makes use of the Hilbert space expanded by all the two-particle 
excited states with the momentum-dependent variational parameters 
to find the best wavefunction.
We demonstrated that the calculated MDF
shows a distinct momentum dependence in contrast to the
Gutzwiller wavefunction and calculated quasi-particle weight shows good
agreement with that obtained by the DMFT combined with the numerical
renormalization group technique (NRG).  The MLA is therefore a new
approach that is competitive
to the DMFT at zero temperature.  Furthermore it allows us to calculate
any static physical quantity because we know the wavefunction itself.

In this letter, we extend the MLA to the first-principles version 
using the LDA+U Hamiltonian.  The first-principles MLA holds high
momentum and total-energy resolutions because of the analytic
formulation and the use of the Laplace
transformation~\cite{pat11,pat13}.  We present the results of the
first-principles MDF bands 
along the high-symmetry lines for bcc Fe for the first time, 
which are definitely not explained by the DFT
or the Gutzwiller wavefunction.
We demonstrate that the MDF for $e_{g}$ and $t_{2g}$ electrons are
strongly momentum-dependent and thus the mass enhancement factors 
calculated from the jump of the MDF are also momentum-dependent.  
The calculated average mass enhancement is consistent with  
the experimental data obtained by the angle-resolved photoemission
spectroscopy (ARPES) as well as those obtained from 
low-temperature specific heat data.  
Our result is also consistent with the recent result of 
LDA+DMFT calculations at finite temperatures, 
though it disagrees with the results of 
the first-principles Gutzwiller theory.

We start from the $spd$ first-principles tight-binding LDA+U Hamiltonian
with intraatomic Coulomb and exchange interactions between $d$ electrons,
and express it as follows assuming a system with one atom per unit cell:
\begin{eqnarray}
H = H_{0} + H_{I} \ .
\label{fphhat}
\end{eqnarray}
The first term $H_{0}$ on the right-hand-side (rhs) is the Hartree-Fock
Hamiltonian.  The second term $H_{I}$ is the residual interaction
given by 
\begin{eqnarray}
H_{I} & = & \sum_{i} 
\Big[ \sum_{L} U_{LL} \, O^{(0)}_{iLL} + \sum_{(L, L^{\prime})} 
(U_{L L^{\prime}} - \frac{1}{2}J_{L L^{\prime}}) 
\, O^{(1)}_{iLL^{\prime}} -
\sum_{(L,L^{\prime})} J_{L L^{\prime}}
O^{(2)}_{iLL^{\prime}}   
\Big] \ . 
\label{fphres}
\end{eqnarray}
Here $L$ denotes the atomic orbitals $lm$ on site $i$, $\sum_{L}$
($\sum_{(L,L^{\prime})}$) expresses the sum over $d$ ($l=2$) orbitals
(pairs between $d$ orbitals), and $U_{LL^{\prime}}$ ($J_{LL^{\prime}}$) is
the intraatomic Coulomb (exchange) integral between electrons on the
orbitals $L$ and $L^{\prime}$.
$O^{(0)}_{iLL}$, $O^{(1)}_{iLL^{\prime}}$, and 
$O^{(2)}_{iLL^{\prime}}$ are the intraorbital operators, 
the interorbital charge-charge operators, and the interorbital 
spin-spin operators, respectively; 
$O^{(0)}_{iLL} = \ \delta\hat{n}_{iL \uparrow} 
\delta\hat{n}_{iL \downarrow}$, 
$O^{(1)}_{iLL^{\prime}} = 
\delta\hat{n}_{iL} \delta\hat{n}_{iL^{\prime}}$, 
$O^{(2)}_{iLL^{\prime}} = \delta\hat{\mbox{\boldmath$s$}}_{iL} \cdot 
\delta\hat{\mbox{\boldmath$s$}}_{iL^{\prime}}$.
Note that $\hat{n}_{iL\sigma}=a^{\dagger}_{iL\sigma}a_{iL\sigma}$, 
$\hat{n}_{iL} = \sum_{\sigma} \hat{n}_{iL\sigma}$, and 
$\hat{\mbox{\boldmath$s$}}_{iL}$
are the electron density operator for an electron with spin $\sigma$ on
site $i$ and orbital $L$, 
the charge density operator, and the spin density operator,
respectively. 
$a^{\dagger}_{iL\sigma}$ ($a_{iL\sigma}$) is the creation
(annihilation) operator for the same electron.
Furthermore $\delta\hat{n}_{iL\sigma}$, for example, should stand for 
$\hat{n}_{iL\sigma} - \langle \hat{n}_{iL\sigma} \rangle_{0}$, 
$\langle \sim \rangle_{0}$ 
being the Hartree-Fock average.

In the first-principles MLA, 
we introduce three momentum-dependent local-ansatz
operators $\tilde{O}^{(\alpha)}_{iLL^{\prime}} \ (\alpha=0, 1, 2)$ 
according to the residual interactions
$O^{(\alpha)}_{iLL^{\prime}}$:
\begin{eqnarray}
\tilde{O}^{(\alpha)}_{iLL^{\prime}} & = & \sum_{ \{ kn\sigma \} } 
\langle k^{\prime}_{2} n^{\prime}_{2}|iL \rangle 
\langle iL|k_{2} n_{2} \rangle
\langle k^{\prime}_{1} n^{\prime}_{1}|iL^{\prime} \rangle 
\langle iL^{\prime}|k_{1} n_{1} \rangle \nonumber \\
& & \hspace{10mm} \times 
\lambda^{(\alpha)}_{LL^{\prime} \{ 2^{\prime}2\,1^{\prime}1 \}} 
\delta(a^{\dagger}_{k^{\prime}_{2}n^{\prime}_{2}\sigma^{\prime}_{2}} 
a_{k_{2}n_{2}\sigma_{2}})
\delta(a^{\dagger}_{k^{\prime}_{1}n^{\prime}_{1}\sigma^{\prime}_{1}} 
a_{k_{1}n_{1}\sigma_{1}}) \ .
\label{fpotilde}
\end{eqnarray}
Here $a^{\dagger}_{kn\sigma}$ ($a_{kn\sigma}$) is the creation
(annihilation) operator for an electron with momentum $\boldsymbol{k}$,  
band index $n$, and spin $\sigma$.  They are given by those in the site
representation as  
$a_{kn\sigma} = \sum_{iL} a_{iL\sigma} \langle kn | iL \rangle$,
$\langle kn | iL \rangle$ being the overlap integrals.

The momentum-dependent amplitudes  
$\lambda^{(\alpha)}_{LL^{\prime} \{ 2^{\prime}2\,1^{\prime}1 \}}$ 
in Eq. (\ref{fpotilde}) are defined by 
\begin{eqnarray}
\lambda^{(0)}_{LL^{\prime} \{ 2^{\prime}21^{\prime}1 \}} & = & 
\eta_{L [ 2^{\prime} 2 1^{\prime} 1 ]}
\delta_{LL^{\prime}} \delta_{\sigma^{\prime}_{2}\downarrow}
\delta_{\sigma_{2}\downarrow} \delta_{\sigma^{\prime}_{1}\uparrow}
\delta_{\sigma_{1}\uparrow} \, ,  \label{fplambda0} \\
\lambda^{(1)}_{LL^{\prime} \{ 2^{\prime}21^{\prime}1 \}} & = & 
\zeta^{(\sigma_{2}\sigma_{1})}_{LL^{\prime} 
[ 2^{\prime} 2 1^{\prime} 1 ] }
\delta_{\sigma^{\prime}_{2}\sigma_{2}}
\delta_{\sigma^{\prime}_{1}\sigma_{1}} \, , \label{fplambda1} \\
\lambda^{(2)}_{LL^{\prime} \{ 2^{\prime}21^{\prime}1 \}} & = & 
\sum_{\sigma} \xi^{(\sigma)}_{LL^{\prime} 
[ 2^{\prime} 2 1^{\prime} 1 ] }
\delta_{\sigma^{\prime}_{2} -\sigma} \delta_{\sigma_{2} \sigma}
\delta_{\sigma^{\prime}_{1}\sigma} 
\delta_{\sigma_{1} -\sigma} + \frac{1}{2} \sigma_{1} \sigma_{2} 
\, \xi^{(\sigma_{2}\sigma_{1})}_{LL^{\prime} 
[ 2^{\prime} 2 1^{\prime} 1 ] } 
\delta_{\sigma^{\prime}_{2}\sigma_{2}}
\delta_{\sigma^{\prime}_{1}\sigma_{1}} \, .  
\label{fplambda2}
\end{eqnarray}
where $\{ 2^{\prime}2\,1^{\prime}1 \}$ 
($[ 2^{\prime}2\,1^{\prime}1 ]$) means    
$\{ k^{\prime}_{2}n^{\prime}_{2}\sigma^{\prime}_{2} 
k_{2}n_{2}\sigma_{2} k^{\prime}_{1}n^{\prime}_{1}\sigma^{\prime}_{1} 
k_{1}n_{1}\sigma_{1} \}$ 
($\{ k^{\prime}_{2}n^{\prime}_{2} 
k_{2}n_{2} k^{\prime}_{1}n^{\prime}_{1} 
k_{1}n_{1} \}$). 
The $\eta$, $\zeta$, and $\xi$ on the rhs are variational
parameters to be determined.

The operators $\tilde{O}^{(0)}_{iLL^{\prime}}$, 
$\tilde{O}^{(1)}_{iLL^{\prime}}$, and 
$\tilde{O}^{(2)}_{iLL^{\prime}}$ describe the intraorbital
correlations, the interorbital charge-charge correlations, and the
interorbital spin-spin correlations, respectively.
Using $\tilde{O}^{(\alpha)}_{iLL^{\prime}}$ and the 
Hartree-Fock wavefunction $|\phi \rangle$, we construct 
the first-principles MLA wavefunction as follows:
\begin{eqnarray}
|\Psi_{\rm MLA} \rangle & = & \Big[ \prod_{i} 
\Big( 1 - \sum_{L} \tilde{O}^{(0)}_{iLL} 
- \sum_{(L, L^{\prime})} \tilde{O}^{(1)}_{iLL^{\prime}} -
\sum_{(L,L^{\prime})} \tilde{O}^{(2)}_{iLL^{\prime}}  \Big) \Big] 
\, |\phi \rangle \ . 
\label{fpmlawf}
\end{eqnarray}

The variational parameters are determined by the stationary condition
for the MLA energy given by
\begin{eqnarray}
E = \langle H \rangle_{0} + N \epsilon_{c} \ . 
\label{fpmlae}
\end{eqnarray}
Here $\langle H \rangle_{0}$ is the Hartree-Fock energy and $N$ is the
number of atoms in the system.  The correlation energy per atom
$\epsilon_{c}$ is obtained in the single-site approximation as 
follows~\cite{kake08}:
\begin{eqnarray}
\epsilon_{\rm c} = \dfrac{-\langle
 \tilde{O}^{\dagger}_{i}\tilde{H}\rangle_{0} -
\langle \tilde{H} \tilde{O}_{i} \rangle_{0} + 
\langle \tilde{O}^{\dagger}_{i}\tilde{H}\tilde{O}_{i}\rangle_{0}}
{1 + \langle \tilde{O}^{\dagger}_{i}\tilde{O}_{i} \rangle_{0}} \ .
\label{mlaec}
\end{eqnarray}
Here $\tilde{O}_{i} = \sum_{\alpha} \sum_{\langle L L^{\prime} \rangle} 
\tilde{O}^{(\alpha)}_{iLL^{\prime}}$ and 
$\tilde{H} = H - \langle H \rangle_{0}$, where 
$\sum_{\langle L L^{\prime} \rangle}$ is defined by a single sum 
$\sum_{L}$ for $L^{\prime} = L$ and a pair sum 
$\sum_{(L, L^{\prime})}$ for $L^{\prime} \neq L$.
Each element in Eq. (\ref{mlaec}) is calculated with use of Wick's
theorem. 

We obtain self-consistent equations for the variational parameters 
$\lambda^{(\alpha)}_{LL^{\prime} \{ 2^{\prime}21^{\prime}1 \}}$ from the
stationary condition $\delta \epsilon_{\rm c} = 0$ and verify that the
solution reduces to the second-order perturbation theory in the weak
Coulomb interaction limit. 
In order to obtain an approximate but explicit solution for a more
correlated regime, we adopt an additional
ansatz to the variational parameters which are exact in the weak
interaction limit~\cite{pat11}:
\begin{eqnarray}
\lambda^{(\alpha)}_{LL^{\prime} \{ 2^{\prime}2\,1^{\prime}1 \}}
= \dfrac{ U^{(\alpha)}_{LL^{\prime}} 
\sum_{\tau} 
C_{\alpha\tau\sigma^{\prime}_{2}\sigma_{2}\sigma^{\prime}_{1}\sigma_{1}}
\tilde{\lambda}^{(\sigma_{2}\sigma_{1})}_{\alpha\tau LL^{\prime}}}
{\epsilon_{k^{\prime}_{2} n^{\prime}_{2} \sigma^{\prime}_{2}} - 
\epsilon_{k_{2} n_{2} \sigma_{2}} - 
\epsilon_{k^{\prime}_{1} n^{\prime}_{1} \sigma^{\prime}_{1}} - 
\epsilon_{k_{1} n_{1} \sigma_{1}}
- \epsilon_{\rm c}} \ .
\label{fplam2}
\end{eqnarray}
Here the Coulomb interaction energy parameters 
$U^{(\alpha)}_{LL^{\prime}}$
are defined as $U_{LL}\delta_{LL^{\prime}} \ (\alpha=0)$,
$U_{LL^{\prime}} - J_{LL^{\prime}}/2 \ (\alpha=1)$,
and $- 2J_{LL^{\prime}} \ (\alpha=2)$.
The coefficients
$C_{\alpha\tau\sigma^{\prime}_{2}\sigma_{2}\sigma^{\prime}_{1}\sigma_{1}}$
are defined by 
$\delta_{LL^{\prime}} \delta_{\sigma^{\prime}_{2}\downarrow}
\delta_{\sigma_{2}\downarrow} \delta_{\sigma^{\prime}_{1}\uparrow}
\delta_{\sigma_{1}\uparrow}$ for $\alpha=0$,
$\delta_{\sigma^{\prime}_{2}\sigma_{2}}
\delta_{\sigma^{\prime}_{1}\sigma_{1}}$ for $\alpha=1$,
$-(1/4) \sigma_{1} \sigma_{2} \delta_{\sigma^{\prime}_{2}\sigma_{2}}
\delta_{\sigma^{\prime}_{1}\sigma_{1}}$
for $\alpha=2, \tau=l$, and  
$-1/2 \sum_{\sigma} 
\delta_{\sigma^{\prime}_{2} -\sigma} \delta_{\sigma_{2} \sigma}
\delta_{\sigma^{\prime}_{1}\sigma} \delta_{\sigma_{1} -\sigma}$ 
for $\alpha=2, \tau=t$.
Note that $l$ ($t$) denotes the longitudinal (transverse) component.
We can regard the renormalization factors
$\tilde{\lambda}^{(\sigma\sigma^{\prime})}_{\alpha\tau LL^{\prime}}$
as new variational parameters.  
The denominator expresses the two-particle excitation energy.
Solving the self-consistent equations obtained by the
variational principle $\delta \epsilon_{\rm c} = 0$, we can determine
these parameters.

We performed the Hartree-Fock band calculations for bcc Fe in
the paramagnetic state and investigated the correlation effects 
using the first-principles MLA.  
We adopted the orbital-independent Coulomb and exchange integrals with 
$U_{LL}=0.2749$ Ry, $U_{LL^{\prime}}=0.1426$ Ry, and
$J_{LL^{\prime}}=0.0662$ Ry 
obtained by Anisimov {\it et al.}~\cite{anis97-2}.
%
%
\begin{figure}
\begin{center}
\includegraphics[scale=0.95]{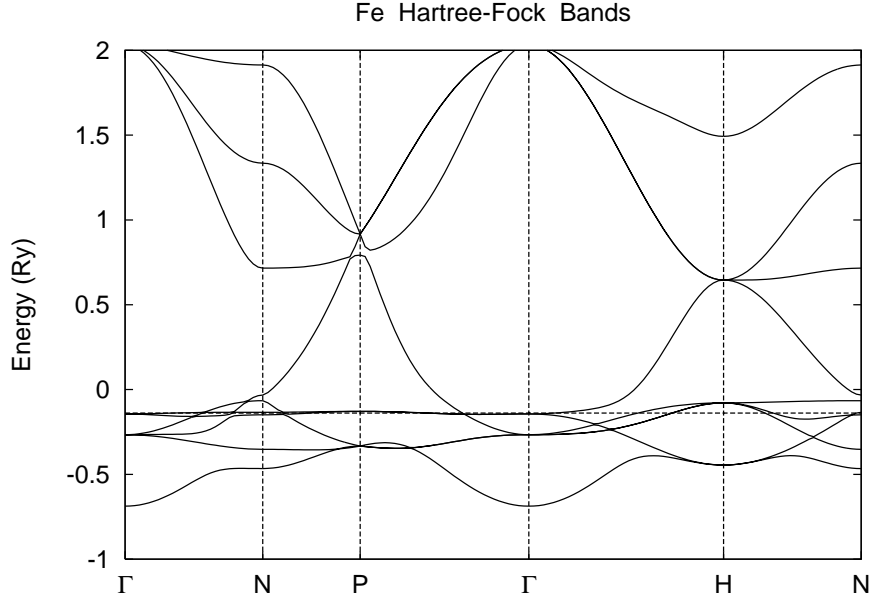}
\end{center}
\caption{Hartree-Fock one-electron energy bands of bcc Fe along 
high-symmetry lines of the first Brillouin zone.  The Fermi level 
($-0.1387$ Ry) is expressed by the horizontal dashed line.
}
\label{fig1}
\end{figure}
%
%

Solving the Hartree-Fock equations for the tight-binding LDA+U
Hamiltonian, we obtained the one-electron energy eigenvalues
$\epsilon_{kn\sigma}$ for paramagnetic Fe.  
Figure 1 shows the energy band curves along 
high-symmetry lines in the first Brillouin zone.  
The band structure for $d$ electrons
in the Hartree-Fock approximation is similar to that obtained by the 
usual LDA band theory.  
Note that the $e_{g}$ bands near the Fermi level 
along the $\Gamma-N-P-\Gamma$ line are
much narrower than the $t_{2g}$ ones.  The other $sp$ bands are mostly
far from the Fermi level ($\epsilon_{F}$);  
thus the Fermi surface of Fe is mainly
determined by the $d$ bands.

The MDF in the first-principles MLA is given by
\begin{eqnarray}
\langle n_{kn\sigma} \rangle  = f(\tilde{\epsilon}_{kn\sigma}) 
+ \dfrac{N \langle \tilde{O}^{\dagger}_{i} \tilde{n}_{kn\sigma} 
\tilde{O}_{i} \rangle_{0}}
{1 + \langle \tilde{O}^{\dagger}_{i}\tilde{O}_{i} \rangle_{0}} \ .
\label{mlank}
\end{eqnarray}
The first term on the rhs is the Fermi-Dirac distribution function 
for the Hartree-Fock independent electrons.  
$\tilde{\epsilon}_{kn\sigma}$ denotes
the energy eigenvalue measured from $\epsilon_{F}$.
The second term is the correlation correction.  Its numerator is
expressed in the paramagnetic state as
\begin{eqnarray}
N \langle \tilde{O}^{\dagger}_{i} \tilde{n}_{kn\sigma} 
\tilde{O}_{i} \rangle_{0} 
= \sum_{\alpha\tau \, \langle L L^{\prime} \rangle} q_{\alpha\tau} 
U^{(\alpha) 2}_{LL^{\prime}} \, 
\tilde{\lambda}^{2}_{\alpha\tau LL^{\prime}} 
\big( \bar{B}_{LL^{\prime}n}(\boldsymbol{k}) 
f(-\tilde{\epsilon}_{kn\sigma}) 
- \bar{C}_{LL^{\prime}n}(\boldsymbol{k}) 
f(\tilde{\epsilon}_{kn\sigma}) \big) \ .
\label{mlank1}
\end{eqnarray}
Here $q_{\alpha\tau}$ is a numerical factor taking the value of  
1 for $\alpha=0$, 2 for $\alpha=1$, 1/8 for $\alpha=2, \tau=l$, 
and 1/4 for $\alpha=2, \tau=t$.
Note that there is no spin dependence in
the variational parameters 
$\tilde{\lambda}_{\alpha\tau LL^{\prime}}$ 
in the paramagnetic state.
$\bar{B}_{LL^{\prime}n}(\boldsymbol{k})$ is a momentum-dependent particle 
contribution above $\epsilon_{F}$ given by
$\bar{B}_{LL^{\prime}n}(\boldsymbol{k}) = |u_{Ln}(\boldsymbol{k})|^{2} 
B_{L^{\prime}Ln}(\epsilon_{kn\sigma}) + 
|u_{L^{\prime}n}(\boldsymbol{k})|^{2} 
B_{LL^{\prime}n}(\epsilon_{kn\sigma})$.  
The hole contribution $\bar{C}_{LL^{\prime}n}(\boldsymbol{k})$ 
is defined by 
$\bar{B}_{LL^{\prime}n}(\boldsymbol{k})$ in which 
$B_{LL^{\prime}n}(\epsilon_{kn\sigma})$ has been replaced by 
$C_{LL^{\prime}n}(\epsilon_{kn\sigma})$.   
The energy-dependent functions $B_{LL^{\prime}n}(\epsilon_{kn\sigma})$
and $C_{LL^{\prime}n}(\epsilon_{kn\sigma})$ are given by the Laplace 
transforms of the local density of states in the Hartree-Fock 
approximations and have been given in the Appendix in Ref. 15
for the single-orbital case.
Note that the MDF in the first-principles MLA
depends on the momentum $\boldsymbol{k}$ not only via the energy
$\epsilon_{kn\sigma}$ but also via $u_{Ln}(\boldsymbol{k})$, 
{\it i.e.}, the eigenvector at each 
$\boldsymbol{k}$ point.
%
%
\begin{figure}
\begin{center}
\includegraphics[scale=1.15]{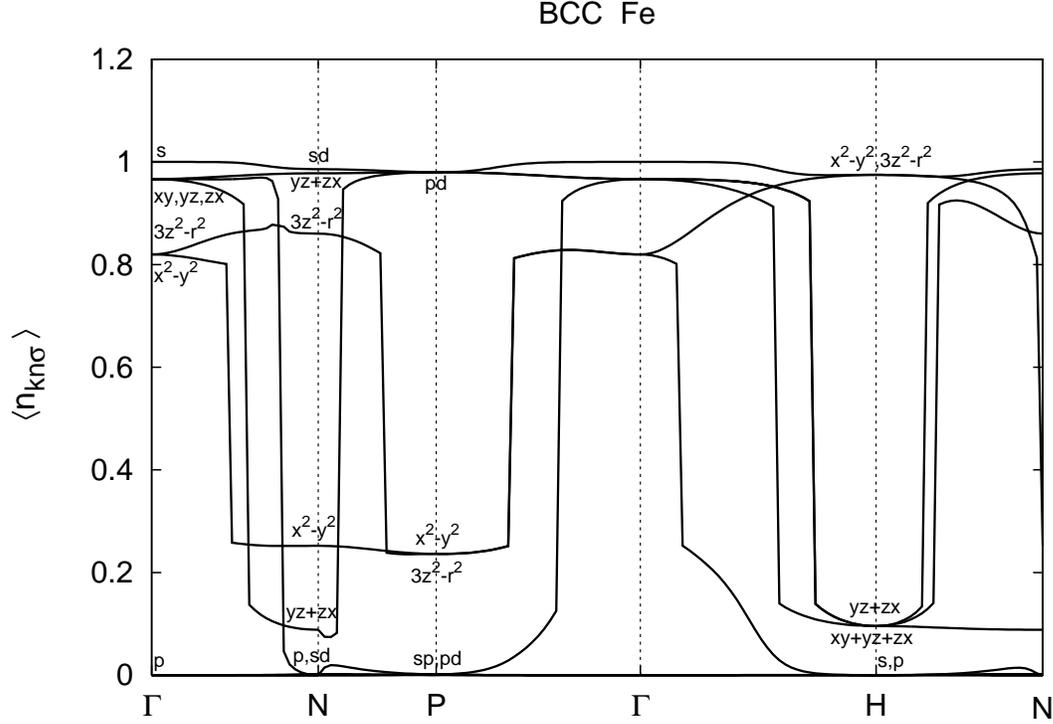}
\end{center}
\caption{Momentum distribution functions 
($\langle n_{kn\sigma} \rangle$) along high-symmetry lines for bcc Fe.  
Orbital symmetry functions and their hybridized states for the branches 
at high-symmetry points are written in the figure. 
}
\label{fig2}
\end{figure}
%
%

We have calculated the MDF in 
paramagnetic Fe using Eq. (\ref{mlank}).  
Figure 2 shows the result along high-symmetry lines.
(Note that the wavevector $\boldsymbol{k}$ is measured in the unit of 
$2\pi/a$, $a$ being the lattice constant.)
We find strong momentum dependence of $\langle n_{kn\sigma} \rangle$
along the lines, which is not described by the Hartree-Fock 
wavefunction.  

At point $\Gamma$, we have a free-electron-like MDF 
$\langle n_{kn\sigma} \rangle = 1.00$ for $s$ electrons with the
Hartree-Fock one-electron energy $\epsilon_{kn\sigma} = -0.69$ Ry 
($< \epsilon_{F}$) (see Fig. 1), while we have the MDF  
$\langle n_{kn\sigma} \rangle = 0.97$ for $d$ electrons with 
$t_{2g}$ symmetry, which are associated with the Hartree-Fock 
one-electron energy $\epsilon_{kn\sigma} = -0.27$ Ry ($< \epsilon_{F}$), 
and the MDF $\langle n_{kn\sigma} \rangle = 0.82$ for $d$ 
electrons with 
$e_{g}$ symmetry with the energy 
$\epsilon_{kn\sigma} = -0.14$ Ry ($< \epsilon_{F}$) in Fig. 1.
For the $p$ electrons associated with the energy 
$\epsilon_{kn\sigma} = 2.03$ Ry ($> \epsilon_{F}$), we again have 
a free-electron-like MDF $\langle n_{kn\sigma} \rangle = 0.00$.

When the momentum $\boldsymbol{k}$ 
moves toward point N along the $\Gamma$-N line,
the MDF for $t_{2g}$ electrons splits into three branches.  The first
branch is almost constant and has a value 
$\langle n_{kn\sigma} \rangle = 0.98$ at point N.
The second branch jumps down at $\boldsymbol{k}_{F} = (0.39, 0.39, 0.00)$ 
on the Fermi surface and approaches 
$\langle n_{kn\sigma} \rangle = 0.00$ at point N.
The third branch decreases with the change in $\boldsymbol{k}$ 
toward point N, 
jumps down at $\boldsymbol{k}_{F} = (0.28, 0.28, 0.00)$, and approaches 
$\langle n_{kn\sigma} \rangle = 0.088$ at point N.
The MDF for $e_{g}$ electrons splits into two branches.  The branch with
$3z^{2}-r^{2}$ symmetry increases and approaches to 
$\langle n_{kn\sigma} \rangle = 0.86$ at point N.  The second branch
with $x^{2}-y^{2}$ symmetry decreases along the $\Gamma$-N lines, jumps
down at $\boldsymbol{k}_{F} = (0.23, 0.23, 0.00)$, and approaches 
$\langle n_{kn\sigma} \rangle = 0.25$ at point N.
The $s$ electron branch of the MDF hardly changes and approaches to the
value $\langle n_{kn\sigma} \rangle = 0.99$ at point N.
The $p$ electron branch also shows flat behavior with 
$\langle n_{kn\sigma} \rangle = 0.00$ 
because there is no hybridization with $d$ electrons and their
one-electron energies are far above the Fermi level (see Fig. 1). 

The basic behavior of the MDF for $s$, $p$, and $d$ electrons mentioned
above is also seen on the other high-symmetry N-P, P-$\Gamma$, $\Gamma$-H,
and H-N lines.  We find that the MDF branches associated with $e_{g}$
electrons show large deviations from 0 and 1, indicating strong electron
correlations. 
The MDF associated with $t_{2g}$ electrons also shows 
significant deviations from 0 and 1.  
On the other hand, the $s$- and $p$-like MDFs
have values close to 1 or 0, indicating that the
independent electron band picture is applicable to their electrons.

The jump of the MDF on the Fermi surface gives the quasi-particle weight
$Z_{kn}$ or the inverse mass enhancement factor 
$(m^{\ast}_{nk}/m)^{-1}$ according to the Fermi liquid theory.  
Since the hybridization between the
$sp$ and $d$ electrons excludes the $sp$-like bands near the Fermi level,
most of the Fermi surface of the bcc Fe is formed by the $d$ bands.
The mass enhancement factors for $e_{g}$ and $t_{2g}$ electrons
calculated along high-symmetry lines are presented in Tables I and II,
respectively.  We find that the mass enhancements for $e_{g}$ electrons
are momentum-dependent and show considerably large values of  
$m^{\ast}_{nk}/m = 1.71 - 1.84$, 
because these electrons form narrow bands
near the Fermi level.  The $t_{2g}$ electrons yield smaller
enhancements of $m^{\ast}_{nk}/m = 1.14 - 1.29$.
\begin{table}[tb]
\caption{Mass enhancement factors for $e_{g}$ electrons at various 
wavevectors $\boldsymbol{k}$ on the Fermi surface.}
\label{tmeffeg}
\begin{tabular}{|c|c|c|c|c|}
\hline
$\boldsymbol{k}$  & (0.23, 0.23, 0.00) & (0.50, 0.50, 0.28) & 
(0.32, 0.32, 0.32) & (0.00, 0.17, 0.00) \\ \hline 
$m^{\ast}_{kn}/m $ & 1.84 & 1.71 & 1.78 & 1.82 \\ \hline
\end{tabular}
\end{table}
\begin{table}[tb]
\caption{Mass enhancement factors for $t_{2g}$ electrons at various 
wavevectors $\boldsymbol{k}$ on the Fermi surface.}
\label{tmefft2g}
\begin{tabular}{|c|c|c|c|c|}
\hline
$\boldsymbol{k}$  & (0.28, 0.28, 0.00) & (0.39,0.39, 0.00) & 
(0.50, 0.50, 0.09) & (0.20, 0.20, 0.20)  \\ \hline 
$m^{\ast}_{kn}/m$ & 1.28 & 1.14 & 1.16 & 1.25  \\ \hline
\end{tabular}
\begin{tabular}{|c|c|c|c|c|}
\hline
$\boldsymbol{k}$  & (0.00, 0.58, 0.00) & (0.00, 0.73, 0.00) & 
(0.15, 0.85, 0.00) & (0.18, 0.82, 0.00) \\ \hline 
$m^{\ast}_{kn}/m$ & 1.29 & 1.27 & 1.27 & 1.29 \\ \hline
\end{tabular}
\end{table}

We have calculated the averaged mass enhancement factor over the Fermi 
surface and obtained $m^{\ast}/m = 1.648$.  
This value is consistent with the
experimental data of $m^{\ast}/m = 1.4 - 2.1$ obtained from the 
low-temperature specific heat~\cite{cheng60,cho96,chio03}.  
Our result also agrees well with the recent result of   
$m^{\ast}/m = 1.7$ obtained by an ARPES experiment~\cite{sanchez09}.
In order to examine the dependence of $m^{\ast}/m$ on the Coulomb
integrals, we performed the same calculations using the alternative set 
$U_{LL}=0.3233$ Ry, $U_{LL^{\prime}}=0.1932$ Ry, and
$J_{LL^{\prime}}=0.0650$ Ry, which was adopted in our LDA+DCPA
calculations~\cite{kake10}.  We obtained $m^{\ast}/m = 1.551$, 
a deviation of 
only 6\% from the value of 1.648. We also suggest that the
ferromagnetic spin polarization may reduce the mass enhancement by about
5\% because of the change in the weight between $e_{g}$ and $t_{2g}$
electrons on the Fermi surface.

The effective mass of Fe has recently been much investigated from the
quantitative viewpoint.  
S\'{a}nchez-Barriga {\it et al.}~\cite{sanchez09} 
performed the three-body theory + LDA-DMFT calculations 
and obtained $m^{\ast}/m = 1.25$ on 
the $\Gamma$-N line, which is too small
as compared with the ARPES result of $m^{\ast}/m = 1.7$.
Katanin {\it et al.}~\cite{katanin10} 
reported the results of 
LDA+DMFT calculations using the quantum Monte-Carlo technique 
(QMC) at 1000 K.  They obtained $m^{\ast}_{t2g}/m = 1.163$ for $t_{2g}$
electrons, in agreement with our result of 
$m^{\ast}_{t2g}/m \approx 1.2$.  However, the value for $e_{g}$ electrons 
was not obtained because of the non-Fermi liquid behavior due to 
strong fluctuations in the narrow $e_{g}$ bands at finite temperatures.
More recently, Pourovskii {\it et al.}~\cite{pour14} 
reported the LDA+DMFT calculations
for bcc Fe using the continuous-time QMC at 300 K.  
They obtained $m^{\ast}/m \approx 1.577$.  The latter is consistent 
with the present result of $m^{\ast}/m = 1.648$.   

First-principles Gutzwiller theory underestimates the mass
enhancement factor.  The LDA+Gutzwiller calculations~\cite{deng09} 
yield a reasonable value of 
$m^{\ast}/m \approx 1.564$, but a too large Coulomb
interaction parameter of $\bar{U}=7.0$ eV was adopted there.  
More recent calculations 
based on the LDA+Gutzwiller theory with reasonable values of 
$\bar{U}=2.5$ eV and $\bar{J}=1.2$ eV 
result in $m^{\ast}_{eg}/m \approx 1.08$ for $e_{g}$ electrons and 
$m^{\ast}_{t2g}/m \approx 1.05$ for $t_{2g}$ electrons~\cite{borghi09}, 
which are too small as compared with the ARPES value of $m^{\ast}/m = 1.7$.

In summary, we have developed a first-principles MLA 
on the basis of the tight-binding
LDA+U Hamiltonian with intra-atomic Coulomb and exchange interactions.
The first-principles MLA reduces to the Rayleigh-Schr\"{o}dinger 
perturbation theory in
the weakly correlated limit, as it should, and quantitatively describes
the correlated electron state with use of the self-consistent 
momentum-dependent variational parameters.

We obtained the first-principles MDF bands for bcc Fe 
along high-symmetry lines, and clarified the band structure of the MDF
for $s$, $p$, and $d$ electrons for the first time.
We found a large deviation from the Fermi-Dirac distribution function for
the branches associated with $e_{g}$ and $t_{2g}$ electrons, while the
$sp$ electron branches follow the usual band theory.
We obtained the momentum-dependent mass enhancement factors 
$m^{\ast}_{kn}/m$ along the high-symmetry line for the first time.
We found the mass enhancements of $m^{\ast}_{eg}/m \approx 1.8$ for 
$e_{g}$ electrons and $m^{\ast}_{t2g}/m \approx 1.2$ for 
$t_{2g}$ electrons, and an average mass enhancement of 
$m^{\ast}/m = 1.648$.  This result explains the recent ARPES value of 
1.7 and the data obtained from the low-temperature specific heat. 
It is also consistent with the theoretical result of 1.577 at 300 K 
based on the recent LDA+DMFT calculations.
First-principles Gutzwiller theory underestimates the mass
enhancement of bcc Fe by 35\% .  For the quantitative description of
bcc Fe, the momentum dependence of the variational parameters is
essential.
The present calculations have been performed in the paramagnetic state.  In
order to discuss more quantitative aspects of the theory, we have to
perform ferromagnetic calculations.  Systematic investigations of
the ferromagnetic Fe, Co, and Ni using the first-principles MLA
are left for future work.
\vspace{3mm}

The present work was supported by a Grant-in-Aid for
Scientific Research (25400404).

\end{document}